\documentclass[oneside]{article}
\usepackage[T1]{fontenc}
\usepackage{authblk}
\usepackage{float}
\usepackage{amsmath}
\usepackage{graphicx}
\usepackage{xfrac}
\usepackage[english]{babel}
\usepackage{lmodern}
\usepackage[T1]{fontenc}
\usepackage[latin9]{inputenc}
\usepackage{mathtools}
\usepackage{graphicx}
\usepackage{esint}
\usepackage[unicode=true,
 bookmarks=false,
breaklinks=false,pdfborder={0 0 1},backref=false,colorlinks=false]
 {hyperref}
\usepackage[a4paper, total={210mm,297mm},margin=2.0cm]{geometry}

\usepackage{datetime}
\newdate{date}{12}{12}{2014}

\providecommand{\tabularnewline}{\\}

\title{Different regimes of the uniaxial elongation of electrically charged
viscoelastic jets due to dissipative air drag}

\author[1]{Marco Lauricella}
\author[1]{Giuseppe Pontrelli \thanks{Electronic address: \texttt{giuseppe.pontrelli@gmail.com}; Corresponding author}}
\author[2]{Ivan Coluzza}
\author[3,4]{Dario Pisignano}
\author[1]{Sauro Succi}

\affil[1]{Istituto per le Applicazioni del Calcolo CNR, Via dei Taurini 19, 00185 Rome, Italy}
\affil[2]{Faculty of Physics, University of Vienna, Boltzmanngasse 5, 1090 Vienna, Austria}
\affil[3]{Dipartimento di Matematica e Fisica Ennio De Giorgi,University of Salento, via Arnesano, 73100 Lecce, Italy}
\affil[4]{Istituto Nanoscienze CNR, Via Arnesano 16, 73100 Lecce, Italy}

\date{\displaydate{date}}

\begin{document}

\maketitle
 
\begin{abstract}
We investigate the effects of dissipative air drag on the dynamics
of electrified jets in the initial stage of the electrospinning process.
The main idea is to use a Brownian noise to model air drag effects
on the uniaxial elongation of the jets. The developed numerical model is used
to probe the dynamics of electrified polymer jets at different conditions
of air drag force, showing that the dynamics
of the charged jet is strongly biased by the presence
of air drag forces. This study provides prospective beneficial implications for
improving forthcoming electrospinning experiments. 
\end{abstract}

\section{Introduction}

In recent years, organic nanofibers have gained a broad fundamental
and industrial interest, due to their peculiar physical properties
and to their numerous potential applications, such as tissue engineering,
air and water filtration, drug delivery and regenerative medicine.
Flexible fibers can be used on a micro- down to nano-scale in electrical,
mechanical and optical systems. In particular, the small cross-section
of electrospun nanofibers in combination with their extreme length
(in principle up to km when polymer solutions with high degree of
molecular entanglement are used) provides a high surface-area ratio
which offers intriguing perspectives for practical applications. As
a consequence, several studies have been focused on the characterization
and production of such one-dimensional (1D) organic nanostructures. Many articles 
\cite{reneker1996nanometre,li2004electrospinning,carroll2008nanofibers,luo2012electrospinning,persano2013industrial,montinaro2015dynamics}
and books \cite{ramakrishna2005introduction,pisignanoelectrospinning,wendorff2012electrospinning}
concerning the production of electrospun nanofibers and the investigation of the phenomenology
of electrified jets have been published in the last
two decades.

\begin{figure}[H]
\begin{centering}
\includegraphics[scale=0.35]{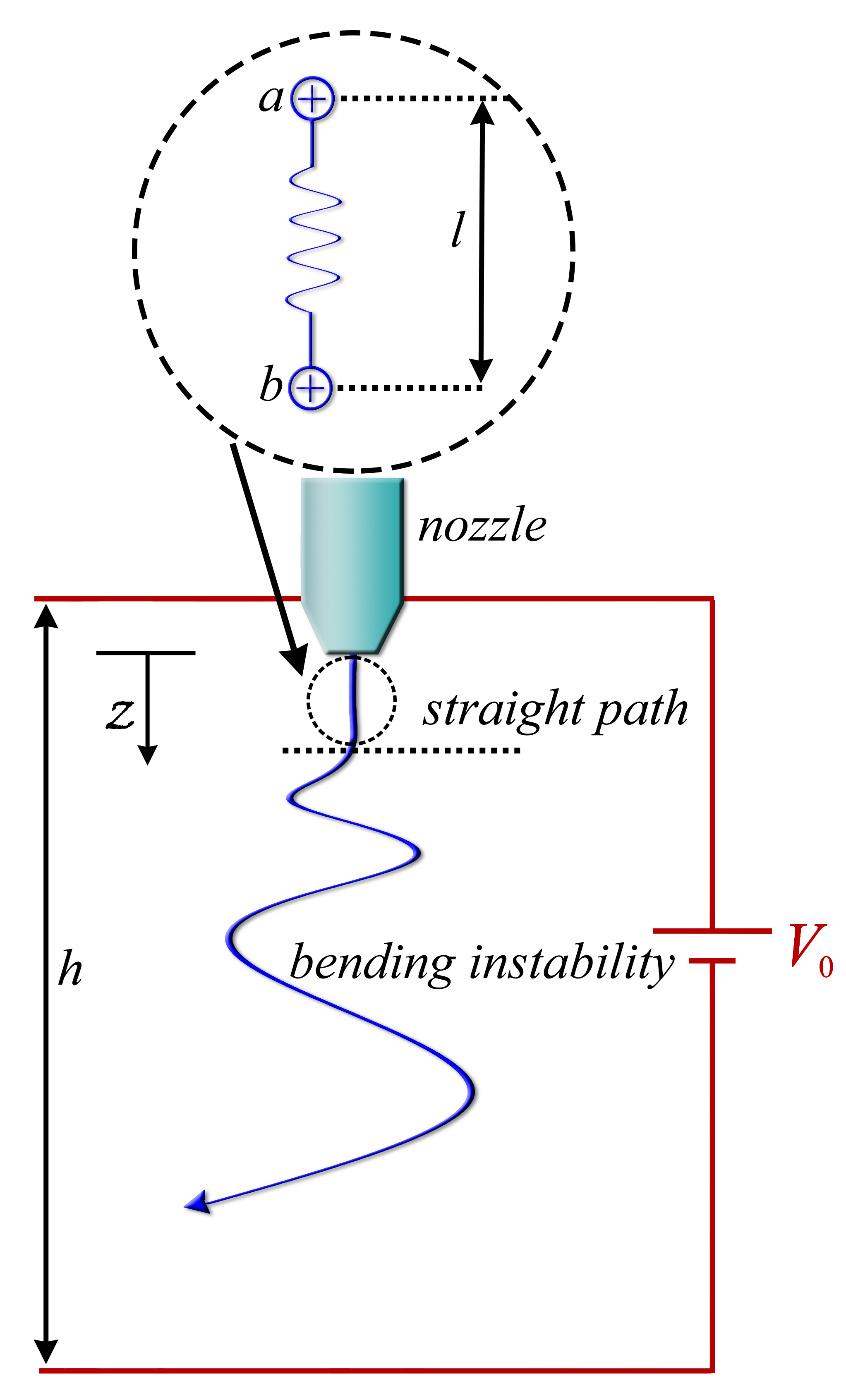} 
\par\end{centering}

\protect\protect\caption{{\small{}ES schematics. $h$: spinneret-collector distance, $V_{0}$: applied voltage bias, $z$: reference axis whose origin is at
the injection point.}}

\label{Fig:schema-esperimento} 
\end{figure}

Following the pioneering works of Rayleigh \cite{rayleigh1882equilibrium}
and, later, Zeleny \cite{zeleny1917instability}, electrospun nanofibers
are synthesized by the uniaxial elongation of
a polymer solution jet, which is ejected at a nozzle from the surface of a charged
droplet (see Fig \ref{Fig:schema-esperimento}). This elongation of the fluid body
is obtained by means of an intense, external electrostatic
field \cite{taylor1964disintegration,taylor1965stability,taylor1969electrically}
(typically $10^{5}-10^{6}\text{ V}\cdot\text{m}^{-1}$) which generates a voltage bias between the
nozzle (spinneret) and a conductive collector. During the jet path from the
nozzle to the collector, the stream cross-section can decrease up
to six orders of magnitude, providing a jet, and consequently solid fibers, with transversal size
well below the micrometer-scale. In a typical electrospinning  
(ES) process the uniaxial elongation of the extruded
polymer jet involves mainly
two sequential stages : 1) an initial quasi-steady stage, 
in which the jet is stretched in a straight path; 2)
a second stage in which relevant bending instabilities might occur, induced by small
perturbations and leading to a jet trajectory spiralling out from the pristine axis of elongation.
The perturbations initiating bending instabilities can be related to mechanical vibrations, as well as to 
hydrodynamic-aerodynamic solicitations along the
jet path. According to the Earnshaw's theorem \cite{jeans1908mathematical},
an off-axis misalignment provides an electrostatic-driven bending instability
before the jet reaches the conductive collector, where the fibers
are finally deposited. As a consequence, the length of the trajectory of the jet increases, and the cross-sectional
size of the elongated fluid body (and of the ultimately deposited nanofibers) undergoes a corresponding further decrease. For these reasons, the design of
electrospinning experiments in which the length of the initial, uniaxial elongation region of the jet is minimized, can be highly desired
when nanofibers with very small diameters are to be produced with a given polymer.

Notwithstanding its major interest, a comprehensive investigation
to understand the transition between different regimes of the jet
dynamics is not complete, and the relation between dissipative-perturbing
forces and the first quasi-steady stage of the ES process
is still in need of further clarification. In particular, an uniaxial
dissipative-perturbing force, such as the air drag force, can reduce the length of 
the initial straight path, so that the overall distance covered by the jet
between the spinneret and the collecting surface increases, and the stream cross-section
is further decreased by subsequent instabilities as mentioned above.

Simulation models can be very useful to rationalize these phenomenona and to 
improve our capability of prediction of which processing parameters mostly affect the fiber morphology.
In the last years, many studies were published in this respect \cite{reneker2000bending,yarin2001taylor,hohman2001electrospinning,fridrikh2003controlling,theron2004experimental,carroll2006electrospinning}. Many of these 
works are based on the equations
of continuum mechanics \cite{spivak2000model,feng2002stretching,feng2003stretching,hohman2001electrospinning,hohman2001applications}. In other studies,
the electrified jet is described as series of discrete elements obeying the equations of Newtonian
mechanics \cite{reneker2000bending,yarin2001taylor}, as in our work.

Here, the dynamics of an electrified polymer
jet is investigated in its early-stage under a stationary dissipative-perturbing force, which is modeled
by a simple Brownian term. Such Brownian noise can efficiently model the stationary
perturbation due to many, simultaneously occurring and tiny impacts along the pristine stretching direction, as those related to air drag
forces affecting the dynamics of the jet while it moves through a gaseous
medium. Such assumption was 
already proposed on the base of reported experiments \cite{antonia1980measurements,ojha2004statistical,ojha2005statistical,sinha2010meltblowing}.
In the present study, we include a dissipative-perturbing force to
the one-dimensional bead-spring model,
developed by Pontrelli \textit{et al.} \cite{pontrelli2014electrospinning},
in order to model the effects
of the air drag force. This is accomplished by adding two force terms
to the set of equations of motion (EOM): a random term and a dissipative
term. These two components of the overall forces obey the fluctuation-dissipation
Langevin relation \cite{reif2009fundamentals}, hence the electrified jet
is described by a Langevin-like stochastic differential equation.
Anticipating the conclusions, we observe that a second quasi steady
stage appears by applying different magnitude of dissipative-perturbing
force. Further, the proposed approach provides a useful starting point
to develop stochastic three-dimensional models of ES
processes.

The article is organized as follows. In Sec. 2 we present our model
for the uniaxial elongation of a viscoelastic jet, and we introduce
the corresponding set of stochastic EOM, which governs the dynamics
of system. Hence, the time integrator of the stochastic EOM is also
reported in Sec. 2. Results are reported and discussed in Sec. 3.
Finally, the Conclusions are outlined in Sec. 4.

\section{Model and time integrator}

We describe a rectilinear jet in the
ES experiment by a viscoelastic dumbbell, $ab$, with two charged beads of
mass $m$, at distance $l$ and having the 
same charge $e$ (Fig. \ref{Fig:schema-esperimento}).
We assume one of the two beads to be fixed, denoted by symbol $a$,
and the other, denoted by $b$, is free to move. A viscoelastic force pulls $b$ back to $a$. 
The collector surface is at a distance $h$ from the spinneret (bead injection point), and a voltage bias $V_{0}$ 
is applied between the two elements. The bead $b$ 
is subject to the force due to the external
electrical field $V_{0}/h$, to the Coulomb repulsive force between the
two beads, and to the dumbbell viscoelastic force. Anticipating the results in
Sec. 4, in this scheme electrostatics and
viscoelasticity compete and determine the first stage of the elongation
process, whereas the second stage of the jet path is mainly governed by the external electrical field.

Overall, the time evolution of the viscoelastic fluid body can be properly described
by the following ordinary differential equations as proposed by Reneker
\textit{\small{}{}et. al.}{\small{}{} \cite{reneker2000bending}
}:

\begin{equation}
m\frac{d\upsilon}{dt}=-\frac{e^{2}}{l^{2}}-\frac{eV_{0}}{h}+\pi r^{2}\sigma,\label{eq:vel-ode}
\end{equation}
where $\upsilon$ is the velocity of the bead $b$, $t$ is time,
$r$ is the cross-sectional radius of the jet filament, $\pi r^{2}\sigma$
is the force pulling the bead $b$ back to $a$ given by the viscoelasticity
of the jet (assumed positive), and $\sigma$ is the stress related to the viscoelastic
force. As evidenced  in Ref. \cite{reneker2000bending}, the 
gravity force is lower than the other force terms involved in Eq. \ref{eq:vel-ode} by many orders of 
magnitude, and it has been neglected. 
Furthermore, we note that in a one-dimensional model, 
the surface tension force 
restoring the rectilinear shape is absent. In addition, for a viscoelastic Maxwellian liquid jet \cite{bird1987dynamics},
the temporal evolution of the $\sigma$  stress is given by:

\begin{equation}
\frac{d\sigma}{dt}=G\frac{dl}{ldt}-\frac{G}{\mu}\sigma,
\end{equation}
where $G$ indicates the elastic modulus, $\mu$ is the viscosity of the
jet solution, and:

\begin{equation}
\frac{dl}{dt}=-\upsilon.
\end{equation}
In order to adopt a non-dimensional form of these equations as is customary
in fluid mechanics \cite{fox2009introduction}, one can use a length
scale $L=\left(e^{2}/\pi r_{0}^{2}G\right)^{1/2}$ with $r_{0}=r\left(t=0\right)$,
and the relaxation time $\tau=\mu/G$. Then we define $\bar{l}=l/L$
in units of the equilibrium length $L$. At $L$, the electrostatic repulsion between beads
matches the reference stress related to viscoelasticity ($G$). The time $\bar{t}=t/\tau$
is the time $t$ rescaled in $\tau$ units. We define $W=-\upsilon$
and $\bar{W}=W\cdot\tau/L$. Applying the condition that the volume
of the jet is conserved, $\pi r^{2}\, l=\pi r_{0}^{2}L$, we write
the set of EOM:

\begin{subequations} 
\begin{alignat}{1}
\frac{d\bar{l}}{d\bar{t}} & =\bar{W}\\
\frac{d\bar{\sigma}}{d\bar{t}} & =\frac{\bar{W}}{\bar{l}}-\bar{\sigma}\\
\frac{d\bar{W}}{d\bar{t}} & =V-F_{ve}\frac{\bar{\sigma}}{\bar{l}}+\frac{Q}{\bar{l}^{2}}
\end{alignat}
\label{eq:ODE} \end{subequations}

where the parameters denoted by bars are dimensionless. The dimensionless
groups are given by:

\begin{subequations} 
\begin{alignat}{1}
Q  & =\frac{e^{2}\mu^{2}}{L^{3}mG^{2}}\\
V  & =\frac{eV_{0}\mu^{2}}{hLmG^{2}}\\
F_{ve}  & =\frac{\pi r_{0}^{2}\mu^{2}}{mLG}
\end{alignat}
\end{subequations}

Overall, the time evolution of the viscoelastic fluid body can be properly described
by the following ordinary differential equations as proposed by Reneker
\textit{\small{}{}et. al.}{\small{}{} \cite{reneker2000bending}
}:

\begin{equation}
m\frac{d\upsilon}{dt}=-\frac{e^{2}}{l^{2}}-\frac{eV_{0}}{h}+\pi r^{2}\sigma,\label{eq:vel-ode}
\end{equation}
where $\upsilon$ is the velocity of the bead $b$, $t$ is time,
$r$ is the cross-sectional radius of the jet filament, $\pi r^{2}\sigma$
is the force pulling the bead $b$ back to $a$ given by the viscoelasticity
of the jet (assumed positive), and $\sigma$ is the stress related to the viscoelastic
force. As evidenced  in Ref. \cite{reneker2000bending}, the 
gravity force is lower than the other force terms involved in Eq. \ref{eq:vel-ode} by many orders of 
magnitude, and it has been neglected. 
Furthermore, we note that in a one-dimensional model, 
the surface tension force 
restoring the rectilinear shape is absent. In addition, for a viscoelastic Maxwellian liquid jet \cite{bird1987dynamics},
the temporal evolution of the $\sigma$  stress is given by:

\begin{equation}
\frac{d\sigma}{dt}=G\frac{dl}{ldt}-\frac{G}{\mu}\sigma,
\end{equation}
where $G$ indicates the elastic modulus, $\mu$ is the viscosity of the
jet solution, and:

\begin{equation}
\frac{dl}{dt}=-\upsilon.
\end{equation}
In order to adopt a non-dimensional form of these equations as is customary
in fluid mechanics \cite{fox2009introduction}, one can use a length
scale $L=\left(e^{2}/\pi r_{0}^{2}G\right)^{1/2}$ with $r_{0}=r\left(t=0\right)$,
and the relaxation time $\tau=\mu/G$. Then we define $\bar{l}=l/L$
in units of the equilibrium length $L$. At $L$, the electrostatic repulsion between beads
matches the reference stress related to viscoelasticity ($G$). The time $\bar{t}=t/\tau$
is the time $t$ rescaled in $\tau$ units. We define $W=-\upsilon$
and $\bar{W}=W\cdot\tau/L$. Applying the condition that the volume
of the jet is conserved, $\pi r^{2}\, l=\pi r_{0}^{2}L$, we write
the set of EOM:

\begin{subequations} 
\begin{alignat}{1}
\frac{d\bar{l}}{d\bar{t}} & =\bar{W}\\
\frac{d\bar{\sigma}}{d\bar{t}} & =\frac{\bar{W}}{\bar{l}}-\bar{\sigma}\\
\frac{d\bar{W}}{d\bar{t}} & =V-F_{ve}\frac{\bar{\sigma}}{\bar{l}}+\frac{Q}{\bar{l}^{2}}
\end{alignat}
\label{eq:ODE} \end{subequations}

where the parameters denoted by bars are dimensionless. The dimensionless
groups are given by:

\begin{subequations} 
\begin{alignat}{1}
Q  & =\frac{e^{2}\mu^{2}}{L^{3}mG^{2}}\\
V  & =\frac{eV_{0}\mu^{2}}{hLmG^{2}}\\
F_{ve}  & =\frac{\pi r_{0}^{2}\mu^{2}}{mLG}
\end{alignat}
\end{subequations}

A reminder for the definitions of the dimensionless parameters is
reported in Table 1. We now extend this model to include air drag effects
on the jet dynamics, adding two further force terms to Eq.
\ref{eq:ODE}c. We indicate as $D_{\upsilon}$
a generic diffusion coefficient in the velocity space and as $\alpha$ a
dissipative term, and we assume that a dissipative force term is present, with form
$\alpha W$. In addition, a random force component is considered, which reads $\sqrt{2D_{\upsilon}}\eta\left(t\right)$,
where $\eta\left(t\right)$ describes a stochastic process that is nowhere
differentiable with $<\eta\left(t_{1}\right)\eta\left(t_{2}\right)>=\delta\left(\left|t_{2}-t_{1}\right|\right)$,
and $<\eta\left(t\right)>=0$. Note that the dissipative term $\alpha W$
is usually dependent on the geometry of the jet, which changes in
time. In particular, based on experimental results, the dissipative
air drag force was proposed equal to $f_{air}=0.65\pi rl\rho_{a}\left(2r/\nu_{a}\right)^{-0.81}W^{1.19}$,
where $\rho_{a}$ and $\nu_{a}$ are the air density and kinematic
viscosity, respectively\cite{spinning1991science}. Assuming a constant
volume of the jet, so that $r=r_{0}\sqrt{L/l}$, we obtain $f_{air}=0.65\pi\rho_{a}r_{0}^{0.19}L^{0.095}l^{0.905}\left(2/\nu_{a}\right)^{-0.81}W^{1.19}$,
provided that the air drag force is depending on the jet length factor
$l^{0.905}$. In this work we assume that $\alpha$ is not $l$-dependent,
supported by the fact that, in the initial stage of the elongation
process, the jet length does not change dramatically. For simplicity, we also model
the dissipative force as a term which increases linearly with $W$
instead of $W^{1.19}$, thus obtaining a simple Langevin-like
stochastic model obeying to the fluctuation-dissipation Langevin relation. 
Thus, adding the two dissipative and randomic force terms in Eq. \ref{eq:vel-ode}, 
we obtain:

\begin{equation} 
m\frac{dW}{dt} = \frac{e^{2}}{l^{2}}+\frac{eV_{0}}{h}-\pi r^{2}\sigma-m\alpha W+\\
+\sqrt{2m^{2}D_{\upsilon}}\eta\left(t\right).
\label{eq:SDE-dimen}
\end{equation}

In order to be consistent with the adopted description, we introduce:

\begin{equation} 
\begin{alignedat}{1}\bar{\alpha} &  =\alpha\tau\\
 \bar{D}_{\upsilon} & =D_{\upsilon}\cdot\tau^{3}/L^{2}
\end{alignedat}
\end{equation}

which are the dimensionless counterparts of $\alpha$ and $D_{\upsilon}$.
Using these definitions and the dimensionless groups, we rewrite the
Eq. \ref{eq:SDE-dimen} as:

\begin{equation}
\frac{d\bar{W}}{d\bar{t}} =V-F_{ve}\frac{\bar{\sigma}}{\bar{l}}+\frac{Q}{\bar{l}^{2}}-\bar{\alpha}\bar{W}+\sqrt{2\bar{D}_{\upsilon}}\eta\left(\bar{t}\right)
.\label{eq:SDE}
\end{equation}
We highlight that $\bar{D}_{\upsilon}$ sets the width of the fluctuations.
In particular, it is possible to demonstrate \cite{gillespie2012simple}
that the variance $\sigma_{\bar{W}}\left(t\right)^{2}$ of the velocity
$\bar{W}$ due to only the random and dissipative terms at time $t$,
computed over an ensemble of stochastic trajectories, is equal to
$\sigma_{\bar{W}}\left(t\right)^{2}\equiv$$<\left[\bar{W}\left(t\right)-<\bar{W}\left(t\right)>\right]^{2}>$
$=\left(\bar{D}_{\upsilon}/\bar{\alpha}\right)\left(1-e^{-2\bar{\alpha}t}\right)$,
and, consequently, $\lim\limits_{t\rightarrow\infty}\sigma_{\bar{W}}\left(t\right)^{2}=\bar{D}_{\upsilon}/\bar{\alpha}$,
which is sometimes called fluctuation-dissipation Langevin relation.

In order to integrate the differential EOM we discretize
$t$ as $t_{i}=t_{0}+i\Delta t$ with $i=1,\ldots,n_{steps}$ where
$n_{steps}$ denotes the number of sub-intervals. In this work we adopt the explicit strong order scheme by Platen \cite{platen1987derivative,kloeden1992numerical,platen2010numerical},
whereof the order of strong convergence was evaluated
equal to 1.5. It is worth to note that for the specific case
under investigation the diffusion coefficient vector has only one
non-zero component equal to $\bar{D}_{\upsilon}$, which is constant.
As consequence, the original integration scheme considerably simplifies.
This scheme avoids the use of derivatives by corresponding finite
differences in the same way as Runge-Kutta schemes do for ODEs in
a deterministic setting, and it is briefly summarized as follows. 

Let us consider a Brownian motion vector process $\textbf{X}=\left\{ \textbf{X}_{t},t\right\} $ of
\textit{d}-dimensional satisfying the stochastic differential equation

\begin{equation}
\frac{d\textbf{X}}{dt}=\textbf{a}\left(t,X^{1},\ldots,X^{d}\right)+\textbf{b}d\Omega\label{eq:generic-sde}
\end{equation}
where $\textbf{a}$ and $\textbf{b}$ are vectors of \textit{d}-dimensional
usually called drift and diffusion vector coefficients, and $\Omega\left(t\right)$
denotes a Wiener process. Denoted $Y_{t}^{k}$ the approximation for
the \textit{k}-th component of the vector $\textbf{X}$ at time $t$, 
the integrator has the following form:

\begin{equation}
Y_{t+\varDelta t}^{k} = Y_{t}^{k}+b^{k}\Delta\Omega+
\frac{1}{2\sqrt{\Delta t}}\left[a^{k}\left(\widetilde{\boldsymbol{\Upsilon}}_{+}\right)-a^{k}\left(\widetilde{\boldsymbol{\Upsilon}}_{-}\right)\right]\Delta\varPsi+ 
 \frac{1}{4}\left[a^{k}\left(\widetilde{\boldsymbol{\Upsilon}}_{+}\right)+2a^{k}+a^{k}\left(\widetilde{\boldsymbol{\Upsilon}}_{-}\right)\right]\Delta t,
\label{eq:integrator-platen}
\end{equation}

with the vector supporting values

\begin{equation} 
\begin{alignedat}{1}\widetilde{\boldsymbol{\Upsilon}}_{\pm} & =\textbf{Y}_{t}+\textbf{a}\Delta t\pm\textbf{b}\sqrt{\Delta t}\\
 \widetilde{\boldsymbol{\Phi}}_{\pm} & =\widetilde{\boldsymbol{\Upsilon}}_{+}\pm\textbf{b}\left(\widetilde{\boldsymbol{\Upsilon}}_{+}\right)\sqrt{\Delta t}.
\end{alignedat}
\end{equation}

Here, $\Delta\Omega$ and $\Delta\varPsi$ indicate normally distributed
random variables constructed from two independent $N\left(0,1\right)$
standard Gaussian distributed random variables ($U_{1}$, $U_{2}$)
by means of the following linear transformation:

\begin{equation} 
\begin{alignedat}{1}\Delta\Omega & =U_{1}\sqrt{\Delta t}\\
 \Delta\varPsi & =\frac{1}{2}\Delta t^{3/2}\left(U_{1}+\frac{1}{\sqrt{3}}U_{2}\right).
\end{alignedat}
\label{eq:random-var-1}
\end{equation}

\section{Results and Discussion}

In the case $\bar{\alpha}=0$ and $\bar{D}_{\upsilon}=0$, Eq. \ref{eq:SDE}
reduces to Eq. \ref{eq:ODE}.c. Consequently, the integration scheme
described by Eq. \ref{eq:integrator-platen} can be used also to integrate
Eqs. \ref{eq:ODE} for the deterministic case. We exploited in preliminary test
the time reversibility to assess a suitable time step $\Delta\bar{t}$, 
which was found equal to $10^{-3}$  in order to provide a precision lower than $10^{-12}$  for the length $\bar{l}$ .

\begin{figure}[H]
\begin{centering}
\includegraphics[scale=0.35]{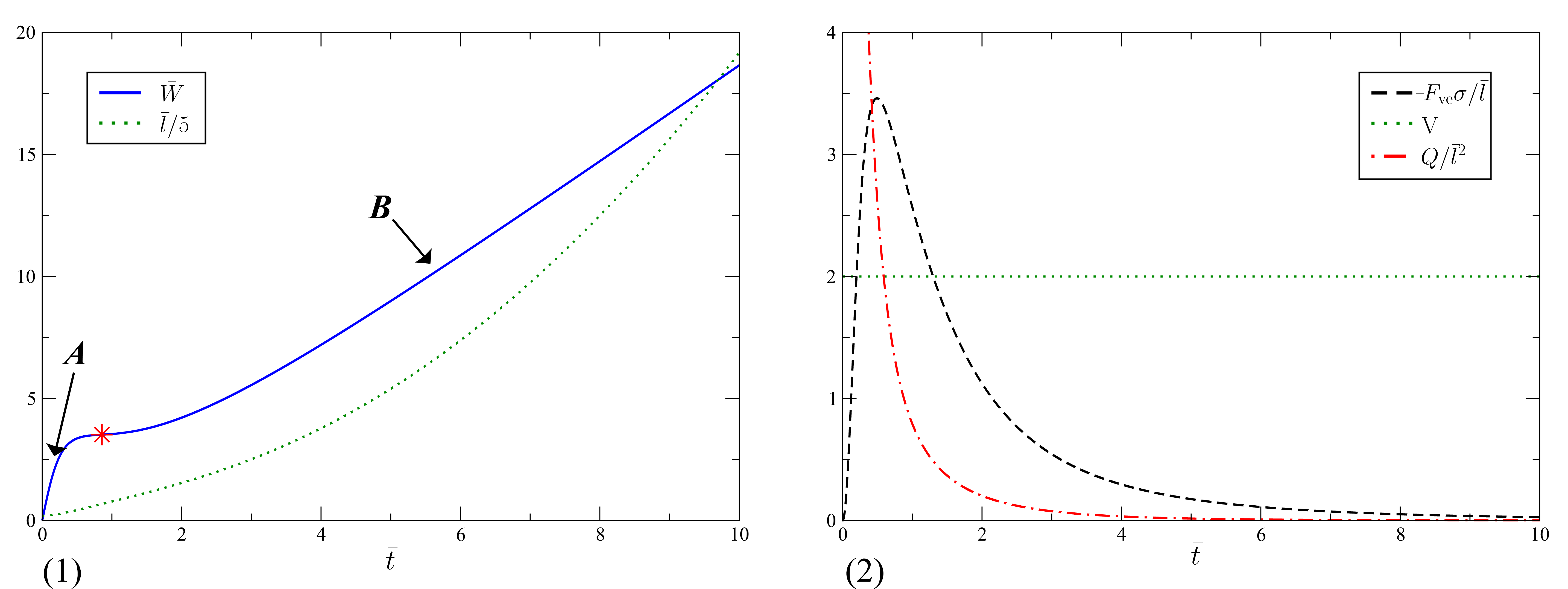} 
\par\end{centering}

\protect\protect\caption{{\small{}Deterministic system. On the left (1) the temporal evolution of 
the length $\bar{l}\left(\bar{t}\right)$ and the velocity $\bar{W}\left(\bar{t}\right)$
(continuous line) with two characteristic regimes (labeled A and B) delimited by 
a quasi stationary point (star). On the right (2) the time evolution of the force terms.}}

\label{Fig:caso-1-l-w} 
\end{figure}

We study and comment on a few, metastable and asymptotic regimes of electrified
jets, associated with the Eq. \ref{eq:SDE}. We consider the typical values $Q=F=12$ and $V=2$ \cite{pontrelli2014electrospinning,reneker2000bending}.
Other parameters, related with 
experiments, are  $L=3.19\,\text{mm}$ and $\tau=
10^{-2}\,\text{s}$ \cite{reneker2000bending}.
The deterministic system case can be described by choosing $\bar{\alpha}=0$
and $\bar{D}_{\upsilon}=0$ . The simulations start with $\bar{l}=1$, $\bar{\sigma}=0$ and $\bar{W}=0$.
In Fig \ref{Fig:caso-1-l-w} we report the time evolution of the velocity
$\bar{W}\left(\bar{t}\right)$. Two sequential stages are seen in the elongation dynamics (point $A$ and $B$ in the Figure). Initially, a moderate increase is found for $\bar{W}\left(\bar{t}\right)$ up to a quasi stationary point ($\bar{t}_{*}$). At this point, corresponding to the lower
limit of the derivative $\partial\bar{W}\left(\bar{t}\right)/\partial\bar{t}$, the viscoelastic force ${\displaystyle \frac{F_{ve}\bar{\sigma}\left(\bar{t}_{*}\right)}{\bar{l}\left(\bar{t}_{*}\right)}}$ equals the sum of ${\displaystyle \frac{Q}{\bar{l}\left(\bar{t}_{*}\right)^{2}}}$
and $V$, thus zeroing the total force. Then,
in the second stage (after $\bar{t}_{*}$) the velocity trend comes to a nearly linearly-increasing
regime, as expected by the relation $\lim\limits_{\bar{t}\rightarrow\infty}\partial\bar{W}\left(\bar{t}\right)/\partial\bar{t}=V$.
Further, we show in Fig \ref{Fig:caso-1-l-w} the time
evolution of the length $\bar{l}$. We observe that $\bar{l}$ increases
as a quadratic term, since the limit of its second derivative is constant,
$\lim\limits_{\bar{t}\rightarrow\infty}\partial\bar{^{2}l}\left(\bar{t}\right)/\partial\bar{t}^{2}=V$.

In Fig. \ref{Fig:caso-1-l-w}
we also report the three force terms ${\displaystyle \frac{Q}{\bar{l}\left(\bar{t}\right)^{2}}}$,
$V$ and ${\displaystyle \frac{F_{ve}\bar{\sigma}\left(\bar{t}\right)}{\bar{l}\left(\bar{t}\right)}}$.
In the first stage we observe an early transient, characterized by
the build-up of the term ${\displaystyle \frac{F_{ve}\bar{\sigma}\left(\bar{t}\right)}{\bar{l}\left(\bar{t}\right)}}$,
which peaks around $\bar{t}=0.5$ under the Coulomb force ${\displaystyle \frac{Q}{\bar{l}\left(\bar{t}\right)^{2}}}$
and the external electric field $V$. In the second stage, the term
${\displaystyle \frac{F_{ve}\bar{\sigma}\left(\bar{t}\right)}{\bar{l}\left(\bar{t}\right)}}$
and ${\displaystyle \frac{Q}{\bar{l}\left(\bar{t}\right)^{2}}}$ start
to decay, and the dynamics tends asymptotically to be governed only
by the term $V$. We point out that a smaller $\bar{t}_{*}$ means a shorter straight path of the electrified jet, so that the overall length of the trajectory covered by the jet from the nozzle and to the collector increases,
and the stream cross-section decreases correspondingly.

We now investigate the jet elongation under stochastic perturbation.
In particular, it is our interest to assess how $\bar{t}_{*}$ is affected the dissipative-perturbing term $-\bar{\alpha}\bar{W}+\sqrt{2\bar{D}_{\upsilon}}\eta\left(\bar{t}\right)$
in Eq. \ref{eq:SDE}. We investigate different magnitudes of $\bar{\alpha}$
keeping the ratio $\bar{D}_{\upsilon}/\bar{\alpha}$ constant. We
stress that the magnitude of the parameters $\bar{\alpha}$ and $\bar{D}_{\upsilon}/\bar{\alpha}$
is depending on the amplitude of the modeled perturbation. As example,
let us consider the aforementioned experimental formula $m\alpha=0.65\pi rl\rho_{a}\left(2r/\nu_{a}\right)^{-0.81}$ in order to assess a typical value of $\bar{\alpha}$.
Taking $\rho_{a}=1210\,\text{ g/m}^{3}$, $\nu_{a}=$ 0.15 $\text{m}^{2}/\text{sec}$,
$r=2\cdot10^{-5}$ m, $l=3.19\cdot10^{-3}$ m (which corresponds to
$l\left(t_{*}\right)$ for the deterministic case) we obtain $m\alpha=7.12\cdot10^{-5}\,\text{g/s}$.
For a typical value of relaxation time $\tau=10^{-2}\,\text{s}$,
and density of the liquid jet $\rho_{l}=1000\,\text{ kg/m}^{3}$,
we obtain a value of $\bar{\alpha}$ equal to 0.18. 

In order to properly
represent the statistical process, 10000 independent trajectories are calculated
for each different value of $\bar{\alpha}$. The
time dependent mean value of physical observables is then computed along the dynamics, and the statistical dispersion is evaluated as interquartile range 
(IQR) \cite{upton1996understanding}. All the trajectories are carried out at the reference values
of $Q=F=12$ and $V=2$.

\begin{figure}[H]
\begin{centering}
\includegraphics[scale=0.60]{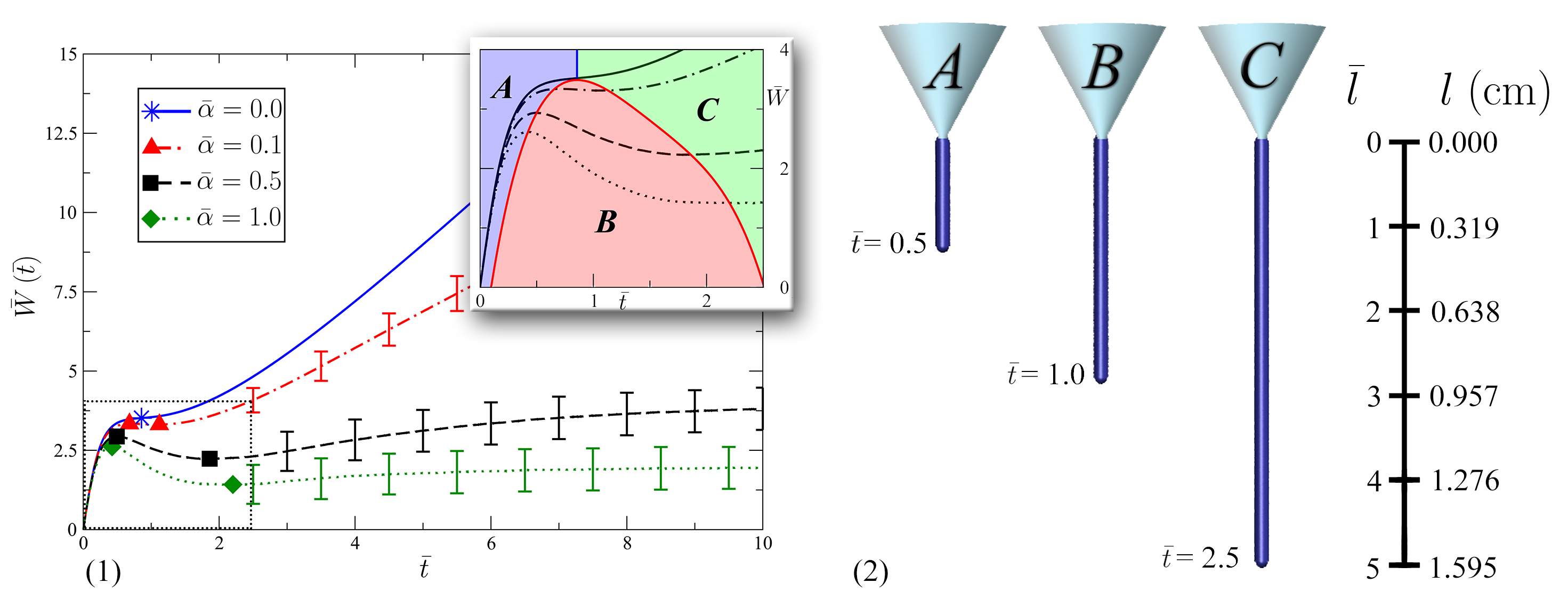} 
\par\end{centering}

\protect\protect\caption{{\small{}On the left (1) the evolution of the velocity $\bar{W}\left(\bar{t}\right)$
for different values of $\bar{\alpha}$ in a stochastic system. Quasi-stationary points are depicted as symbols
for all the $\bar{\alpha}$ values. The error bars are computed as
IQR. In the rectangular inset the initial trend $\bar{W}\left(\bar{t}\right)$ is enlarged,
and the three sequential stages of the uniaxial elongation process
are labeled $A$ (blue), $B$ (red) and $C$ (green). On the right (2) three snapshots of the ES simulation with $\bar{\alpha}=1.0$, in
scale, taken at three different times $\bar{t}$, each one corresponding to a specific stage.}}

\label{Fig:caso-stoc-v} 
\end{figure}

\begin{figure}[H]
\begin{centering}
\includegraphics[scale=0.35]{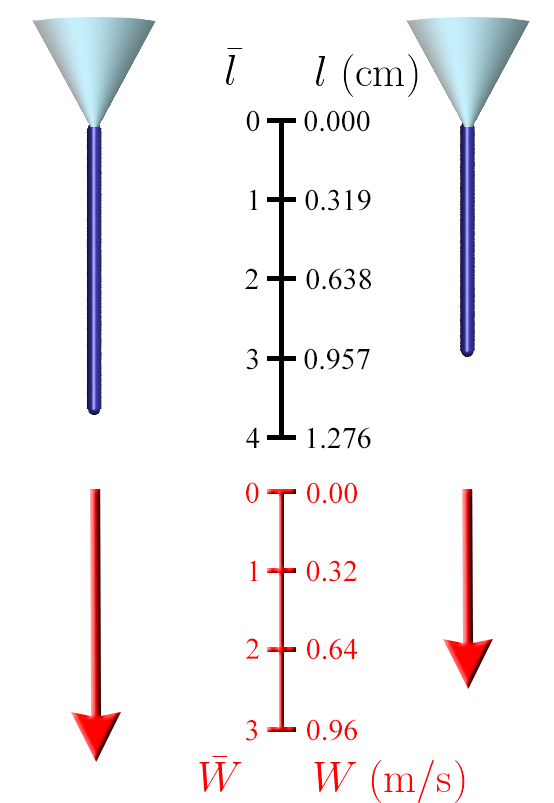} 
\par\end{centering}

\protect\protect\caption{{\small{}Two snapshots of the ES simulations, in
scale, taken at time $\bar{t}=1.0$ ($t=0.01\: \text{s}$), showing the jet elongation for two different cases:
the deterministic case (left), and the stochastic case with $\bar{\alpha}=1.0$
(right). The velocities $\bar{W}$ are also drawn (bottom) as red vectors.
Both dimensional and non-dimensional quantities are reported.
}}

\label{Fig:snapshot-conf} 
\end{figure}

In Fig. \ref{Fig:caso-stoc-v} we show the time evolution of the velocity
$\bar{W}\left(\bar{t}\right)$ for three values of the friction
coefficient $\bar{\alpha}$. First of all, we note, in all the cases
with $\bar{\alpha}\neq0$ the presence of two quasi-stationary points
denoted $\bar{t}_{*}$ and $\bar{t}_{**}$, instead of only one, like
in the deterministic system. Here, $\bar{t}_{*}$ is the point of
coordinates $\left(\bar{t}_{*},\bar{W}\left(\bar{t}_{*}\right)\right)$,
where the system reaches the condition $\partial\bar{W}\left(\bar{t}\right)/\partial\bar{t}=0$
for the first time, and similarly $\bar{t}_{**}$ is the point $\left(\bar{t}_{**},\bar{W}\left(\bar{t}_{**}\right)\right)$
where the system reaches the condition $\partial\bar{W}\left(\bar{t}\right)/\partial\bar{t}=0$
for the second time. 
This second point $\bar{t}_{**}$
is originated by the dissipation air drag term,
which slows down the dynamics of the jet. 
This effecy is evident in Fig. \ref{Fig:snapshot-conf},
where the jet dynamics in the presence and in the absence of the air 
drag force are compared.
Interestingly, since a slower dynamics makes the fluid body more exposed to bending instabilities, the air drag effect can likely 
increase the ultimate jet elongation and lead to a further reduction 
of the filament cross-section, 
which plays in favour of the transversal miniaturization of the finally collected nanofibers. We point out that, although the present 1D model is not able to describe bending instabilities, it represents a preliminary
step for developing three-dimensional stochastic models of ES processes.

By using the points $\bar{t}_{*}$ and $\bar{t}_{**}$ we now define
three sequential stages of the uniaxial elongation process (denoted
in Figure \ref{Fig:caso-stoc-v} as \textit{A}, \textit{B} and C, respectively). 

\begin{figure}[H]
\begin{centering}
\includegraphics[scale=0.35]{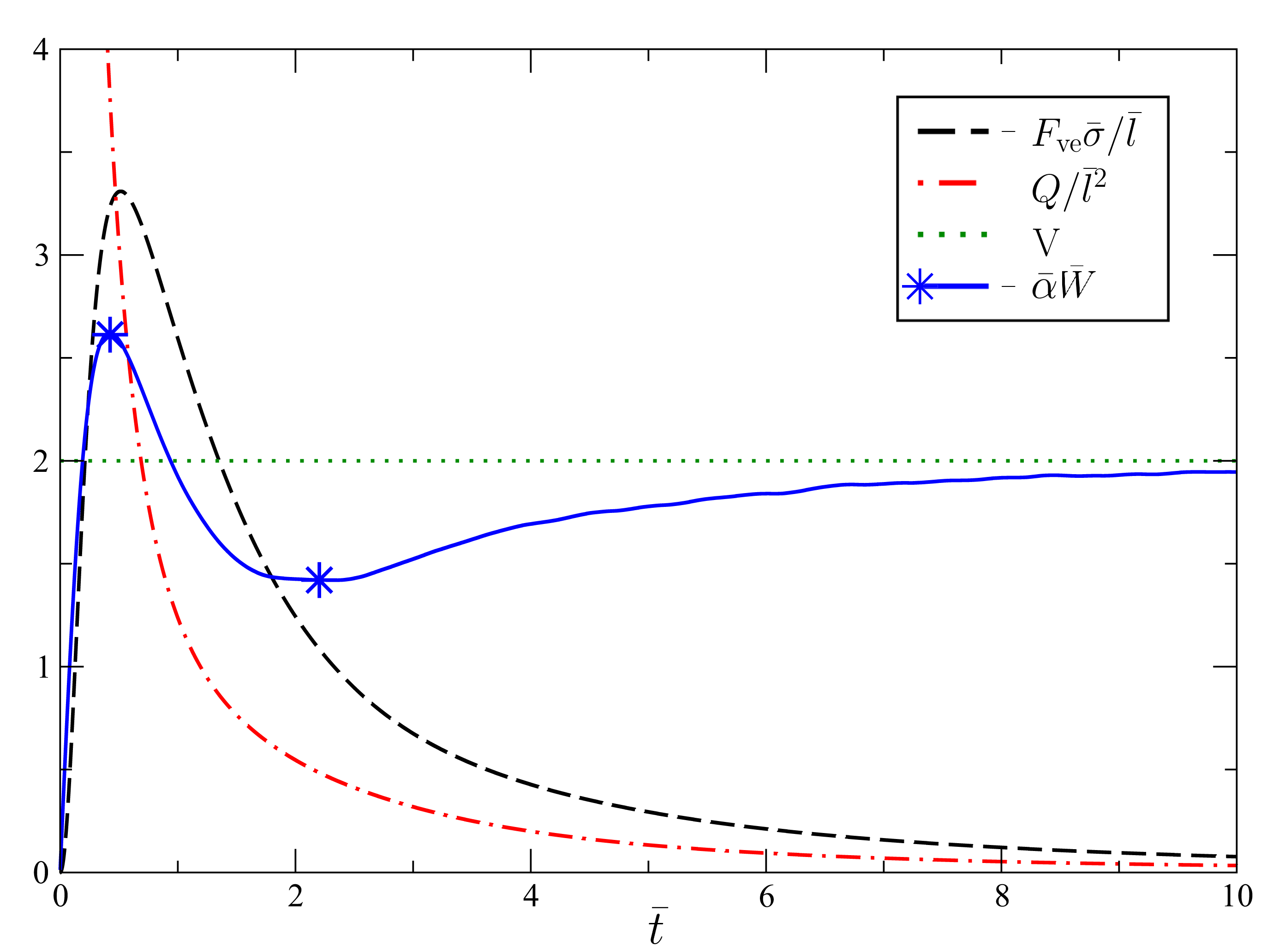} 
\par\end{centering}

\protect\protect\caption{{\small{}Stochastic system. Evolution of the force terms. The quasi-stationary points
are depicted as star symbols. The term $-\bar{\alpha}\bar{W}$ tends asymptotically to a stationary
regime.}}

\label{Fig:stoc-force} 
\end{figure}

In order to analyze the different regimes, we examine the
force terms reported in Figure \ref{Fig:stoc-force} for the case $\bar{\alpha}=1$. 
First of all, we observe that the Coulomb term ${\displaystyle \frac{Q}{\bar{l}\left(\bar{t}\right)^{2}}}$
decays rapidly, playing a secondary role. 

The first stage \textit{A}
is characterized by the ${\displaystyle \frac{F_{ve}\bar{\sigma}\left(\bar{t}\right)}{\bar{l}\left(\bar{t}\right)}}$
and $\bar{\alpha}\bar{W}$ terms, increasing due to jet stretching as induced by the external field $V$. 
The combined action of the opposite forces 
produces the first quasi stationary point $\bar{t}_{*}$, where the
terms ${\displaystyle \frac{F_{ve}\bar{\sigma}\left(\bar{t}\right)}{\bar{l}\left(\bar{t}\right)}}$
and $\bar{\alpha}\bar{W}$ balance the term $V$. 
In the second stage \textit{B}, the terms ${\displaystyle \frac{F_{ve}\bar{\sigma}\left(\bar{t}\right)}{\bar{l}\left(\bar{t}\right)}}$
and $\bar{\alpha}\bar{W}$ become larger in modulus than the opposite
electrostatic term ($V$), so that we observe a decrease of the velocity $\bar{W}$. 
At the same time, the term ${\displaystyle \frac{F_{ve}\bar{\sigma}\left(\bar{t}\right)}{\bar{l}\left(\bar{t}\right)}}$
starts to decay, and the sum of the two terms ${\displaystyle \frac{F_{ve}\bar{\sigma}\left(\bar{t}\right)}{\bar{l}\left(\bar{t}\right)}}$
and $\bar{\alpha}\bar{W}$ becomes insufficient to balance the opposite
Coulomb term $V$. Thus, we observe the second quasi stationary point
$\bar{t}_{**}$.

\begin{figure}[H]
\begin{centering}
\includegraphics[scale=0.35]{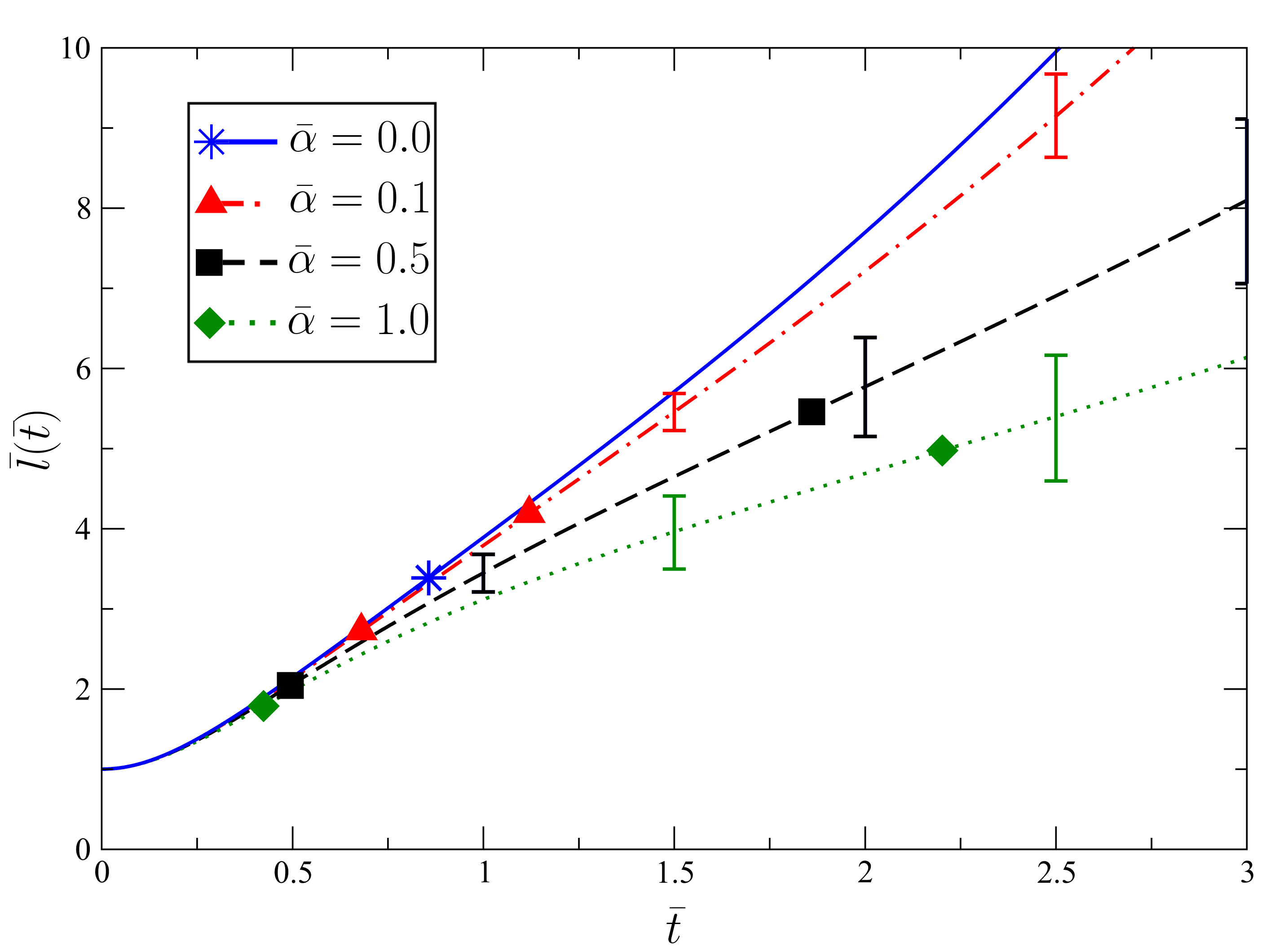} 
\par\end{centering}

\protect\protect\caption{{\small{}Evolution of $\bar{l}\left(\bar{t}\right)$
for different values of $\bar{\alpha}$ equal to 0 (continuous line),
0.1 (dashed-dotted line), 0.5 (dashed line), and 1 (dotted line). $\bar{D}_{\upsilon}/\bar{\alpha}=1$. Quasi-stationary points are indicated as in Fig. \ref{Fig:caso-stoc-v}.
The error bars are computed as IQR.}}

\label{Fig:caso-stoc-l} 
\end{figure}

In the third stage, the jet
dynamics is finally governed by the remaining opposite terms $V$ and
$\bar{\alpha}\bar{W}$, since the term ${\displaystyle \frac{F_{ve}\bar{\sigma}\left(\bar{t}\right)}{\bar{l}\left(\bar{t}\right)}}$ tends to zero. 
As consequence, the velocity $\bar{W}$ rises again
under the external electrical force $V$, to achieve asymptotically
a final stationary regime, where the dissipative force $\bar{\alpha}\bar{W}$
balances completely $V$. In Table \ref{tab:PQSvalues} the
values of $\bar{t}_{*}$ and $\bar{t}_{**}$ are summarized for all the three values of
$\bar{\alpha}$. We note that the gap between $\bar{t}_{*}$ and $\bar{t}_{**}$
becomes larger by increasing $\bar{\alpha}$, and the path length
through the second stage increases (see Figure \ref{Fig:caso-stoc-v}).
Note that the straight path of the electrified jet is described by
the observable $\bar{l}\left(\bar{t}_{*}\right)$. We observe a decrease
of the length $\bar{l}\left(\bar{t}_{*}\right)$ upon increasing $\bar{\alpha}$
(Figure \ref{Fig:caso-stoc-l}), namely the length covered by the jet during its initial elongational stage 
is reduced due to the early uniaxial
perturbation. 
It is worth pointing out that the linear Langevin model employed in this work leads to an underestimate of the drag effect, which is generally reported 
to be mildly superlinear.
Even though this non-linearity may lead to significant effects on the long-term dynamics of the system, it is unlikely to affect the short-term one, which is the focus of the present paper. 
The non-linear friction is being studied in forthcoming work, and surely warrants a thorough investigation for the case of fully three-dimensional long-term dynamics.

Advantages for ES processes coming from
unveiling and characterizing these effects can be numerous,
including the possibility of better controlling the dynamics of electrified
jets. In particular, the diameter of collected nanostructures
could be significantly reduced by exploiting air drag effects, which can increase the jet path 
by slowing down the dynamics of the process.

\section{Conclusions}

Electrified viscoelastic fluid bodies, as those typical of ES experiments, were analyzed
under conditions comprising stationary stochastic perturbations. A Brownian
term has been employed to model a stationary dissipative-perturbing
force. The resulting effects on the jet stretching were investigated. The main finding is that perturbation
forces, such as air drag force, change significantly the ES dynamics, leading to the presence of a second
quasi stationary point. Further, we find that the jet linear
extension in the early ES stage decreases upon increasing dissipation. 
These conclusions may allow experimental conditions embedding increased dissipative components to be designed, which might enhance
the efficiency of the ES process and the capability to produce ultra-thin polymer fibers. 
Examples of such conditions may include gas flows in the process
atmosphere, and mechanical solicitations resulting in enhanced ambient vibrations.

\section*{Acknowledgments}

This research has been funded by the European Research Council under
the European Unions Seventh Framework Programme (FP/2007-2013)/ERC
Grant Agreement n. 306357 (ERC Starting Grant NANO-JETS).

\newpage

\section*{Tables}

\begin{table}[H]
\begin{centering}
\begin{tabular}{cccc}
\hline 
Physical  & Dimensional  & Dimensionless  & Dimensionless\tabularnewline
Parameter  & Symbol (units)  & Symbol  & Definition\tabularnewline
\hline 
\hline 
Time  & $t$ $\left(\text{s}\right)$  & $\bar{t}$  & $t/\tau$\tabularnewline
Length of the rectilinear jet part  & $l$ $\left(\text{m}\right)$  & $\bar{l}$  & $l/L$\tabularnewline
Velocity  & $\upsilon$ $\left(\text{m/s}\right)$  & $\bar{\upsilon}$  & $\upsilon\cdot\tau/L$\tabularnewline
Absolute velocity  & $W$ $\left(\text{m/s}\right)$  & $\bar{W}$  & $W\cdot\tau/L$\tabularnewline
Stress  & $\sigma$ $\left(\text{g/(cm }\text{s}^{2}\text{)}\right)$  & $\bar{\sigma}$  & $\sigma/G$\tabularnewline
Friction coefficient  & $\alpha$ $\left(\text{s}^{-1}\right)$  & $\bar{\alpha}$  & $\alpha\tau$\tabularnewline
Velocity diffusion coefficient  & $D_{\upsilon}$ $\left(\text{m}^{2}\text{/s}^{3}\right)$  & $\bar{D}_{\upsilon}$  & $D_{\upsilon}\cdot\tau^{3}/L^{2}$\tabularnewline
\hline 
\end{tabular}
\par\end{centering}

{\small{}\protect\protect\caption{{\small{}Definitions of the dimensionless parameters employed in the
text.  We remind that $G$ is the elastic modulus,
$\mu$ the viscosity of jet, $\tau=\mu/G$ the relaxation time, and
$L$ the equilibrium length at which Coulomb repulsion matches the
reference viscoelastic stress $G$.}}
}{\small \par}

\label{Tab:definitions-of-parameters} 
\end{table}

\begin{table}[H]
\begin{centering}
\begin{tabular}{cccccccccc}
\hline 
$\bar{\alpha}$  & $\bar{D}_{\upsilon}/\bar{\alpha}$  & $\bar{t}_{*}$  & $\bar{l}\left(\bar{t}_{*}\right)$  & $\bar{\sigma}\left(\bar{t}_{*}\right)$  & $\bar{W}\left(\bar{t}_{*}\right)$  & $\bar{t}_{**}$  & $\bar{l}\left(\bar{t}_{**}\right)$  & $\bar{\sigma}\left(\bar{t}_{**}\right)$  & $\bar{W}\left(\bar{t}_{**}\right)$\tabularnewline
\hline 
\hline 
0  & 0  & 0.86  & 3.39  & 0.81  & 3.52  & ...  & ...  & ...  & ...\tabularnewline
0.1  & 1  & 0.68  & 2.73  & 0.74  & 3.34  & 1.12  & 4.19  & 0.82  & 3.31\tabularnewline
0.5  & 1  & 0.49  & 2.04  & 0.58  & 2.94  & 1.86  & 5.46  & 0.60  & 2.23\tabularnewline
1  & 1  & 0.42  & 1.79  & 0.48  & 2.61  & 2.20  & 4.98  & 0.44  & 1.42\tabularnewline
\hline 
\end{tabular}
\par\end{centering}

\protect\protect\caption{{\small{}Values of the adimensional variables: length $\bar{l}$, stress
$\bar{\sigma}$ and velocity $\bar{W}$ at the points
$\bar{t}_{*}$ and $\bar{t}_{**}$, for different values of
$\bar{\alpha}$. For the deterministic case we report the data of
the unique quasi-stationary point $\bar{t}_{*}$. A larger gap is found between $\bar{t}_{*}$ and $\bar{t}_{**}$ upon increasing
$\bar{\alpha}$, as well as a lower $\bar{l}\left(\bar{t}_{*}\right)$ (Fig. \ref{Fig:caso-stoc-l}),
which indicates that the initial elongation stage is reduced by the uniaxial perturbation.}}

\label{tab:PQSvalues} 
\end{table}

\newpage

\section*{Figures}

\begin{thebibliography}{10}
\expandafter\ifx\csname url\endcsname\relax
  \def\url#1{\texttt{#1}}\fi
\expandafter\ifx\csname urlprefix\endcsname\relax\def\urlprefix{URL }\fi
\expandafter\ifx\csname href\endcsname\relax
  \def\href#1#2{#2} \def\path#1{#1}\fi

\bibitem{reneker1996nanometre}
D.~H. Reneker, I.~Chun, Nanometre diameter fibres of polymer, produced by
  electrospinning, Nanotechnology 7~(3) (1996) 216.

\bibitem{li2004electrospinning}
D.~Li, Y.~Wang, Y.~Xia, Electrospinning nanofibers as uniaxially aligned arrays
  and layer-by-layer stacked films, Advanced Materials 16~(4) (2004) 361--366.

\bibitem{carroll2008nanofibers}
C.~P. Carroll, E.~Zhmayev, V.~Kalra, Y.~L. Joo, Nanofibers from electrically
  driven viscoelastic jets: modeling and experiments, Korea-Aust Rheol J 20
  (2008) 153--164.

\bibitem{luo2012electrospinning}
C.~Luo, S.~D. Stoyanov, E.~Stride, E.~Pelan, M.~Edirisinghe, Electrospinning
  versus fibre production methods: from specifics to technological convergence,
  Chemical Society Reviews 41~(13) (2012) 4708--4735.

\bibitem{persano2013industrial}
L.~Persano, A.~Camposeo, C.~Tekmen, D.~Pisignano, Industrial upscaling of
  electrospinning and applications of polymer nanofibers: a review,
  Macromolecular Materials and Engineering 298~(5) (2013) 504--520.

\bibitem{montinaro2015dynamics}
M.~Montinaro, V.~Fasano, M.~Moffa, A.~Camposeo, L.~Persano, M.~Lauricella,
  S.~Succi, D.~Pisignano, Sub-ms dynamics of the instability onset of
  electrospinning, Soft Matter 11 (2015) 3424--3431.

\bibitem{ramakrishna2005introduction}
S.~Ramakrishna, K.~Fujihara, W.-E. Teo, T.-C. Lim, Z.~Ma, An introduction to
  electrospinning and nanofibers, Vol.~90, World Scientific, 2005.

\bibitem{pisignanoelectrospinning}
D.~Pisignano, Polymer Nanofibers: Building Blocks for Nanotechnology, Royal
  Society of Chemistry, 2013.

\bibitem{wendorff2012electrospinning}
J.~H. Wendorff, S.~Agarwal, A.~Greiner, Electrospinning: materials, processing,
  and applications, John Wiley \& Sons, 2012.

\bibitem{rayleigh1882equilibrium}
L.~Rayleigh, On the equilibrium of liquid conducting masses charged with
  electricity, Philosophical Magazine Series 5 14~(87) (1882) 184--186.

\bibitem{zeleny1917instability}
J.~Zeleny, Instability of electrified liquid surfaces, Phys. Rev. 10 (1917)
  1--6.

\bibitem{taylor1964disintegration}
G.~Taylor, Disintegration of water drops in an electric field, Proceedings of
  the Royal Society of London. Series A. Mathematical and Physical Sciences
  280~(1382) (1964) 383--397.

\bibitem{taylor1965stability}
G.~Taylor, A.~McEwan, The stability of a horizontal fluid interface in a
  vertical electric field, Journal of Fluid Mechanics 22~(01) (1965) 1--15.

\bibitem{taylor1969electrically}
G.~Taylor, Electrically driven jets, Proceedings of the Royal Society of
  London. A. Mathematical and Physical Sciences 313~(1515) (1969) 453--475.

\bibitem{jeans1908mathematical}
J.~H. Jeans, The mathematical theory of electricity and magnetism, University
  Press, 1908.

\bibitem{reneker2000bending}
D.~H. Reneker, A.~L. Yarin, H.~Fong, S.~Koombhongse, Bending instability of
  electrically charged liquid jets of polymer solutions in electrospinning,
  Journal of Applied physics 87~(9) (2000) 4531--4547.

\bibitem{yarin2001taylor}
A.~L. Yarin, S.~Koombhongse, D.~H. Reneker, Taylor cone and jetting from liquid
  droplets in electrospinning of nanofibers, Journal of Applied Physics 90~(9)
  (2001) 4836--4846.

\bibitem{hohman2001electrospinning}
M.~M. Hohman, M.~Shin, G.~Rutledge, M.~P. Brenner, Electrospinning and
  electrically forced jets. i. stability theory, Physics of Fluids 13~(8)
  (2001) 2201--2220.

\bibitem{fridrikh2003controlling}
S.~V. Fridrikh, H.~Y. Jian, M.~P. Brenner, G.~C. Rutledge, Controlling the
  fiber diameter during electrospinning, Physical review letters 90~(14) (2003)
  144502.

\bibitem{theron2004experimental}
S.~Theron, E.~Zussman, A.~Yarin, Experimental investigation of the governing
  parameters in the electrospinning of polymer solutions, Polymer 45~(6) (2004)
  2017--2030.

\bibitem{carroll2006electrospinning}
C.~P. Carroll, Y.~L. Joo, Electrospinning of viscoelastic boger fluids:
  Modeling and experiments, Physics of Fluids 18~(5) (2006) 053102.

\bibitem{spivak2000model}
A.~Spivak, Y.~Dzenis, D.~Reneker, A model of steady state jet in the
  electrospinning process, Mechanics research communications 27~(1) (2000)
  37--42.

\bibitem{feng2002stretching}
J.~Feng, The stretching of an electrified non-newtonian jet: A model for
  electrospinning, Physics of Fluids 14~(11) (2002) 3912--3926.

\bibitem{feng2003stretching}
J.~Feng, Stretching of a straight electrically charged viscoelastic jet,
  Journal of Non-Newtonian Fluid Mechanics 116~(1) (2003) 55--70.

\bibitem{hohman2001applications}
M.~M. Hohman, M.~Shin, G.~Rutledge, M.~P. Brenner, Electrospinning and
  electrically forced jets. ii. applications, Physics of Fluids 13~(8) (2001)
  2221--2236.

\bibitem{antonia1980measurements}
R.~Antonia, B.~Satyaprakash, A.~Hussain, Measurements of dissipation rate and
  some other characteristics of turbulent plane and circular jets, Physics of
  Fluids (1958-1988) 23~(4) (1980) 695--700.

\bibitem{ojha2004statistical}
R.~Ojha, P.-A. Lemieux, P.~Dixon, A.~Liu, D.~Durian, Statistical mechanics of a
  gas-fluidized particle, Nature 427~(6974) (2004) 521--523.

\bibitem{ojha2005statistical}
R.~Ojha, A.~Abate, D.~Durian, Statistical characterization of the forces on
  spheres in an upflow of air, Physical Review E 71~(1) (2005) 016313.

\bibitem{sinha2010meltblowing}
S.~Sinha-Ray, A.~L. Yarin, B.~Pourdeyhimi, Meltblowing: I-basic physical
  mechanisms and threadline model, Journal of Applied Physics 108~(3) (2010)
  034912.

\bibitem{pontrelli2014electrospinning}
G.~Pontrelli, D.~Gentili, I.~Coluzza, D.~Pisignano, S.~Succi, Effects of
  non-linear rheology on the electrospinning process: a model study, Mechanics
  Research Communications 61 (2014) 41--46.

\bibitem{reif2009fundamentals}
F.~Reif, Fundamentals of statistical and thermal physics, Waveland Press, 2009.

\bibitem{bird1987dynamics}
R.~B. Bird, R.~C. Armstrong, O.~Hassager, Dynamics of polymeric liquids. vol.
  1: Fluid mechanics.

\bibitem{fox2009introduction}
W.~Fox~Robert, T.~McDonald~Alan, J.~Pritchard~Philip, Introduction to fluid
  mechanics (2009).

\bibitem{spinning1991science}
A.~Ziabicki, H.~Kawai, High-Speed Fiber Spinning: Science and Engineering
  Aspects, Krieger Publishing Co, 1991.

\bibitem{gillespie2012simple}
D.~T. Gillespie, E.~Seitaridou, Simple Brownian Diffusion: An Introduction to
  the Standard Theoretical Models, Oxford University Press, 2012.

\bibitem{platen1987derivative}
E.~Platen, Derivative free numerical methods for stochastic differential
  equations, in: Stochastic Differential Systems, Springer, 1987, pp. 187--193.

\bibitem{kloeden1992numerical}
P.~E. Kloeden, E.~Platen, Numerical solution of stochastic differential
  equations, Vol.~23, Springer, 1992.

\bibitem{platen2010numerical}
E.~Platen, N.~Bruti-Liberati, Numerical solution of stochastic differential
  equations with jumps in finance, Vol.~64, Springer, 2010.

\bibitem{upton1996understanding}
G.~Upton, I.~Cook, Understanding statistics, Oxford University Press, 1996.

\end{thebibliography}
\end{document}